\documentstyle[12pt]{article}
\def\be {\begin{equation}}
\def\ee {\end{equation}}
\def\ba {\begin{eqnarray}}
\def\ea {\end{eqnarray}}

\def\bi {\begin{itemize}}
\def\ei {\end{itemize}}
\begin{document}
\def\bea{\begin{eqnarray}}
\def\eea{\end{eqnarray}}
\title{\bf {Large-$N$ limit of the  two-dimensinal Non-Local Yang-Mills theory
on arbitrary surfaces with boundary }}
 \author{M.R. Setare  \footnote{E-mail: rezakord@ipm.ir}
  \\{Physics Dept. Inst. for Studies in Theo. Physics and
Mathematics(IPM)}\\
{P. O. Box 19395-5531, Tehran, IRAN }}
\date{\small{}}

\maketitle
\begin{abstract}
 The large-$N$ limit of the two-dimensional non-local
U$(N)$ Yang-Mills theory on an orientable  and non-orientable
surface with boundaries is studied. For the case which the
holonomies of the gauge group on the boundaries are near the
identity, $U\simeq I$, it is shown that the phase structure of
these theories  is the same as that obtain for these theories on
orientable  and non-orientable surface without boundaries, with
same genus but with a modified area $V+\hat{A}$.
 \end{abstract}
\newpage

 \section{Introduction}
The two-dimensional Yang-Mills theory (YM$_{2}$) is a theoretical
laboratory for testing ideas and concepts about four-dimensional
QCD and also string theory. Moreover this theory was studied at
large N beginning with \cite{3} from a number of viewpoints
\cite{{4},{5},{6},{7},{8},{9}}. This theory has an exact stringy
description in the limit of large number N of colours
\cite{{3},{4},{6},{7}}. It was shown that the coefficients of the
$1/N$ expansion of the partition function of SU$(N)$ YM$_2$ are
determined by a sum over maps from a two-dimensional surface onto
the two-dimensional target space.\\
Recently Vafa \cite{vafa} has shown that topological strings on a
class of non-compact Calabi-Yau threefolds is equivalent to two
dimensional bosonic U$(N)$ Yang-Mills on a torus. This
correspondence come from the recent result on the equivalenc of
the partition function of topological strings and that of four
dimensional BPS black holes \cite{osv} \be Z_{BH}=|Z_{top}|^{2}
\ee On the other hand the black hole partition function is given
by the partition function of a quantum field theory living on the
brane which produces the black hole \be
Z_{brane}=|Z_{top}|^{2}=Z_{BH} \ee The partition function of the
field theory on the brane reduces, for the ground state sector, to
the YM$_2$ theory. In \cite{{vafa},{2}} the proposal of \cite{osv}
was made more concrete by considering Calabi-Yau back ground of
the form \be L_1\oplus L_2\longrightarrow \Sigma_{g} \ee where
$\Sigma_{g}$ is a Riemann surface of genus $g$ and $L_1, L_2$ are
line bundles such that $deg(L_1)+deg(L_2)=2g-2$. In this case, the
relevant brane gauge theory reduces to a q-deformed version of
two-dimensional YM theory on the Riemann surface $\Sigma_{g}$.
q-deformed YM$_2$ can be regarded as a one-parameter deformation
of the standard two-dimensional Yang-Mills theory \cite{ro} (to
see recent progress in this topic refer to \cite{{seb},{hola},
{11},{25}}). In this paper we do not study these important and
very interesting progress.  Only I would like to
emphasis on the new progress has been made in YM$_2$ theory.\\
It is well known that YM$_2$ is defined by the Lagrangian
$tr(\frac{1}{4}F^2)$ on a Riemann surface where $F$ is the field
strength tensor. This theory have certain properties, such as
invariance under area preserving diffeomorphism and lack of any
propagating degrees of freedom \cite{b1}. In a YM$_2$ one starts
from a B-F theory in which a Lagrangian of the form $i{\rm
tr}(BF) + {\rm tr}(B^2)$ is used where $B$ is an auxiliary field
in the adjoint representation of the gauge group. There are,
however, the many way to generalized these theories without losing
properties. One way is so- called non-local YM$_2$ (nlYM$_2$,)and
that is to use a non-local action for the auxiliary field
\cite{kh1}. It is remarkable that,  the action of nlYM$_2$ is no
extensive. In non-local YM$_2$ theories, the solution  appear as
some infinite summations over the irreducible representations of
the gauge group. In the large - N limit, however, these
summations are replaced by suitable path integrals over
continuous parameters characterizing the Young tableaux, and
saddle-point analysis shows that the only significant
representation is the classical one, which minimizes some
effective action. This continuous parameters characterizing the
representation  is a constrained, as the length of the rows of
the Young tableau is non-increasing. So for small values of the
surface area, the classical solution satisfies the constraint;
for large values of the surface area, it dose not. Therefore  the
dominating representation is not  the one, which minimizes the
effective  action. This introduces a phase transition  between
these  two regime.\\In this paper we would like  study the
non-local two dimensional $U(N)$ Yang-Mills (nlYM$_2$) theories
on an  arbitrary orientable and non-orientable surface with
boundaries. It is interesting to test the conjecture (2) for
these types of theories, we hope to come back at future to this
important problem.
\section{Preliminaries}
The partition function of a nlYM$_2$ on an orientable
two-dimensional surface $\Sigma_{g,n}$ with genus $g$ and $n$
boundaries is as
\begin{equation}\label{z}
 Z_{g,n}(U_1,\dots,U_n;A)=\sum_Rd_R^{2-2g-n}\chi_R(U_1)
\cdots\chi_R(U_n)\,e^{\omega [-A \Lambda(B)]}.
\end{equation}
$A$ is the surface area, $R$'s label the irreducible
representation of the gauge group,  $d_R$ is the dimension of the
representation $R$, and $\chi_{R}(U)$ is the character of the
holonomy $U_j$. $\omega$ is an arbitrary function of $B$, where
$B$ is an auxiliary field in the adjoint representation of the
gauge group. $\Lambda(B)$ is given in term of the representation
$R$ as following
\begin{equation}\label{lam}
\Lambda=\sum_{k=1}^{p}\frac{\alpha_{k}}{N^{k-1}}C_{k}(R).
\end{equation}
Here $C_{k}$ is the $k$th Casimir of gauge group, $\alpha_{k}$'s
are arbitrary constant. As one can see from (\ref{z}),
corresponding to each boundary a factor $\chi_R(U_i)/d_R$ appears
in the expression for
the partition function.\\
The representation $R$ of the gauge group U$(N)$ is characterized
by $N$ integers $l_1$ to $l_N$, satisfying \cite{khor}
\begin{equation}
+\infty >l_1>l_2>\cdots >l_N>-\infty .
\end{equation}
 The $k$th Casimir of gauge group is given by
\begin{equation}\label{casi}
 C_{k}(R)=\sum_{i=1}^{N}[l_{i}^{k}-(N-i)^{k}].
\end{equation}
The group element $U$ has $N$ eigenvalues $s_1=e^{i\theta_1}$ to
$s_N=e^{i\theta_N}$. The character $\chi_R(U)$ is then
\begin{equation}\label{ch}
  \chi_R(U)=\frac{\det\left\{e^{i l_j\theta_k}\right\}}
          {\mathrm{van}(s_1,\dots ,s_N)},
\end{equation}
where $\mathrm{van}(s_1,\dots ,s_N)$ is the van der Monde
determinant
\begin{equation}
\mathrm{van}(s_1,\dots ,s_N)= \prod_{i<j} (s_i-s_j).
\end{equation}
If we expand $\chi_R(U_i)/d_R$ around $U\approx I$ for the group
$U(N)$ we obtain (for more details see \cite{khor})
\begin{equation}\label{26}
\ln\left[\frac{\chi_R(U)}{d_R}\right]=a\sum_i(s_i-1)+b\sum_i(s_i-1)^2+c\Bigg[\sum_i
(s_i-1)\Bigg]^2+ \cdots,
\end{equation}
where
\begin{equation}\label{aeq}
a=\frac{1}{N}\sum_il_i-\frac{N-1}{2},
\end{equation}
\begin{equation}\label{beq}
b=\frac{1}{2(N^2-1)}\Bigg(\sum_i l_i^2\Bigg)
-\frac{1}{2N(N^2-1)}\Bigg(\sum_i l_i\Bigg)^2
-\frac{1}{2N}\Bigg(\sum_i l_i\Bigg)+\frac{5N-6}{24},
\end{equation}
\begin{equation}\label{ceq}
 c=
-\frac{1}{2N(N^2-1)}\Bigg(\sum_il_i^2\Bigg)+\frac{1}{2N^2(N^2-1)}
\Bigg(\sum_i l_i\Bigg)^2+\frac{1}{24}.
\end{equation}
\section{The large-$N$ limit of the U$(N)$ partition function}
In the large-$N$ limit, one introduces the continuous variables
\cite{rus}
\begin{equation}\label{con}
\phi (x)=-\frac{l_{i}(x)}{N},\hspace{1cm}0\leq x=\frac{i}{N}\leq
1,
\end{equation}
which represent the irreducible representation. In the large-$N$
limit,
\begin{equation}
\sum_i f(l_i)\to N\int_0^1\mathrm{d}x\;f[-N\phi (x)].
\end{equation}
So,
\begin{equation}\label{a1eq}
a=-N\left[ \int_0^1\mathrm{d}x\;\phi(x)+\frac{1}{2} \right],
\end{equation}
\begin{equation}\label{b1eq}
b=\frac{N}{2}\left\{\int_0^1\mathrm{d}x\;\phi^2(x)
-\left[\int_0^1\mathrm{d}x\;\phi(x)\right]^2
+\int_0^1\mathrm{d}x\;\phi(x)+\frac{5}{12}\right\},
\end{equation}
\begin{equation}\label{c1eq}
c=\frac{1}{2}\left\{-\int_0^1\mathrm{d}x\;\phi^2(x)
+\left[\int_0^1\mathrm{d}x\;\phi(x)\right]^2+\frac{1}{12}
\right\}.
\end{equation}
In the large-$N$ limit, the discrete eigenvalues
$s_j=e^{i\theta_j}$ are also represented by the eigenvalue density
function $\sigma (\theta )$ with $\theta \in [-\pi ,\pi ]$, and
one has \cite{15}
\begin{equation}
\sum_i f(\theta_i)\to
N\int_{-\pi}^{\pi}\mathrm{d}\theta\;\sigma(\theta)\, f(\theta ).
\end{equation}
So, if $U\approx I$ we have
\begin{equation}
s_j-1=i\theta_j-\theta_j^2/2+\cdots,
\end{equation}
Inserting Eqs.(\ref{a1eq}, \ref{b1eq}, \ref{c1eq}) in (\ref{26}),
one obtain
\begin{equation}\label{S1}
\ln\left[\frac{\chi_R(U)}{d_R}\right]=-iN^2Q(U)\left[\int\mathrm{d}x\,\phi(x)+\frac{1}{2}\right]
-\frac{N^2}{2}V(U)\left\{\int\mathrm{d}x\;\phi^2(x)-\left[\int\mathrm{d}x\;\phi(x)\right]^2
-\frac{1}{12}\right\}
\end{equation}
where
\begin{equation}\label{q}
Q(U)=\int\mathrm{d}\theta\;\sigma(\theta)\,\theta
\end{equation}
\begin{equation}\label{u}
V(U)=\int\mathrm{d}\theta\;\sigma(\theta)\,\theta^2
-\left[\int\mathrm{d}\theta\;\sigma(\theta)\,\theta\right]^2,
\end{equation}
For the remaining part of the partition function one has
\begin{equation}
d_R^\eta\, e^{\omega [-A \Lambda(B)]}=e^{S_0},\qquad\eta=2-2g,
\end{equation}
and \cite{seta}( Note that $l_i$ in this paper is the same as
$n_i-i+N$ in \cite{seta}).
\begin{equation}\label{seq}
S_0[\phi ] = -N^{2}\Omega{\Biggr (}A \int _{0}^{1} W[\phi(x)]
dx{\Biggl )} +N^{2} (1-g)\int_{0}^{1} dx \int_0^{1} dy
\log|\phi(x)- \phi(y)|,
\end{equation}
where \be W[\phi(x)]= \sum_k (-1)^k \alpha_{k}  \phi^k(x). \ee
Here we have redefined the function $\omega$ as
\begin{equation}\label{redf}
\omega(-A \Lambda(R))=-N^2 \Omega(A
\sum_{k=1}^{p}\alpha_{k}\hat{C_{k}}(R))
\end{equation}
where
\begin{equation}\label{casi1}
\hat{C_{k}}(R)=\frac{1}{N^{k+1}}\sum_{i=1}^{N}l_{i}^{k}
\end{equation}
 The large-$N$ limit of the
partition function (\ref{z}) then becomes the following functional
integral
\begin{equation}\label{39}
Z_{g,n}(U_1,\cdots,U_n;A)=\int{\cal D}\phi\;e^{S[\phi]},
\end{equation}
where
\begin{equation}
S[\phi ]=S_0[\phi]+S'[\phi],
\end{equation}
in which
\begin{equation}\label{41}
S'[\phi]=-iN^2\,Q\left[\int\mathrm{d}x\;\phi(x)+\frac{1}{2}\right]
-\frac{N^2}{2}\,V\left\{\int\mathrm{d}x\;\phi^2(x)-
\left[\int\mathrm{d}x\;\phi(x)\right]^2-\frac{1}{12}\right\},
\end{equation}
where
\begin{equation}\label{42}
Q=\sum_j Q(U_j)\hspace{1cm} V=\sum_j V(U_j).
\end{equation}
The phase structure of the action $S_0[\phi]$ is equivalent to
the phase structure  of the following action \cite{seta}
\begin{equation}\label{13}
S'_0[\phi ] = -N^2\hat{A} \int _{0}^{1} W[\phi(x)] dx +N^2
(1-g)\int_{0}^{1} dx \int_0^{1} dy \log|\phi(x)- \phi(y)|,
\end{equation}
with
 \be\label{14} \hat{A} = 2A \Omega'{\Biggr (}A \int _{0}^{1}
W[\phi(x)] dx{\Biggl )}. \ee
Defining
\begin{equation}
q=\int_0^1\mathrm{d}x\;\left[\phi(x)+\frac{1}{2}\right],\hspace{1cm}
\psi(x)=\phi(x)+\frac{1}{2}-q,
\end{equation}
one arrives at
\begin{equation}
Z_{g,n}(U_1,\cdots,U_n;A)=Z_1\,Z_2,
\end{equation}
where
\begin{equation}
Z_1 =\exp\left(-\frac{N^2\,Q^2}{2A}\right),
\end{equation}
and
\begin{equation}\label{Z}
Z_2=e^{N^2V/24}\int{\cal D}\psi\;\exp\Bigg\{-\frac{N^2}{2}
\Bigg[2\hat{A}\int\mathrm{d}x\;W[\psi(x)]+V\int dx \psi^{2}(x)
-\eta\int\mathrm{d}x\;\mathrm{d}y\;\log|\psi(x)-\psi(y)|\Bigg]\Bigg\}.
\end{equation}
For the special case $W[\psi(x)]=\frac{\psi^{2}}{2}$ we have
\begin{equation}\label{Zm}
Z_2=e^{N^2V/24}\int{\cal D}\psi\;\exp\Bigg\{-\frac{N^2}{2}
\Bigg[(\hat{A}+V)\int dx \psi^{2}(x)
-\eta\int\mathrm{d}x\;\mathrm{d}y\;\log|\psi(x)-\psi(y)|\Bigg]\Bigg\}.
\end{equation}
 It is seen that the partition function (\ref{Zm}) is in fact equal to
the partition function on a closed orientable surface with genus
$g$ and modified surface $\hat{A}+V$: \be
Z_2=\hat{Z}_{g,0}(V+\hat{A}) \ee Then \be\label{45}
\log[Z_{g,n}(U_1,\cdots,U_n;A)]=\frac{N^2}{2}\left(\frac{V}{12}-\frac{Q^2}{A}\right)
+\log[\hat{Z}_{g,0}(\hat{A}+V)].
\end{equation}
As one can see, the logarithm of the partition function on a
surface with boundaries has been written in terms of the
logarithm of the partition function on a surface without
boundaries, but with modified area \footnote{This does not modify
the actual area of the surface, of course. The point is that
$\log[Z_{g,0}(A)]$ is known to have a discontinuity (in its third
derivative) at $A=A_{c}$. From Eq.(\ref{45}), it is seen that
$\log(Z)$ has a discontinuity in its third derivative, at
$\hat{A}+V=A_{c}$. This means that the critical area is $A_c-V$.
So there is a phase transition only for $\Sigma_{0,n}$, and the
transition occures at
\begin{equation}\label{ac}
A_{\mathrm{c}}=\pi^2-V.
\end{equation}}.
This relation is an approximate and is valid only if the
holonomies corresponding to the boundaries don't differ much from
unity. The area-dependence of the partition-function
corresponding to a surface without boundaries is known, and it is
known that at some specific area ($\pi$ squared for ordinary
Yang-Mills\cite{dk}) it undergoes a third order transition, that
is, at this point the third derivative of the logarithm of the
partition function is discontinuous. The remaning term is
obviously analytic, so it does not change the discontinuity of
the third derivative. Hence the phase structure of the non-local
two-dimensional Yang-Mills theory on an orientable surface with
boundary come from $\hat{Z_{g,0}}(V+\hat{A})$ and it has third
order phase
transition only on surfaces with $g=0$.\\
Now we consider the similar problem for two-dimensional non-local
Yang-Mills theory on a non-orientable surface with area $A$,
genus $g$, $s$ copies of Klein bottle, $r$ copies of projective
plan, and $n$ boundary. The partition function of this theory is
as
\begin{equation}\label{zn}
 Z_{g,n}(U_1,\dots,U_n;A)=\sum_Rd_R^{2-(2g+2s+r+n)}\chi_R(U_1)
\cdots\chi_R(U_n)\,e^{\omega [-A \Lambda(B)]}.
\end{equation}
where the summation is only over self-conjugate representation of
the gauge group. This requirement in $U(N)$ means that there is
the additional constraint to the sums as \be n_i=-n_{N-i+1} \ee
In the large-N limit, this implies that the continuum variables,
$\phi(x)$, satisfy \be \phi(x)=-\phi(1-x) \ee So one can define a
new function such as \be \phi(x)=\psi(x)\hspace{0.3cm}
for\hspace{0.5cm} 0\leq x \leq 1/2;
\hspace{1cm}\phi(x)=-\psi(1-x)\hspace{0.3cm} for \hspace{0.5cm}
1/2 \leq x \leq 1 \ee Here the function $\psi(x)$ being defined
on the interval $[0,1/2]$, in which $\psi(1/2)=0$. Then, by
applying this constraint to the large-N limit of (\ref{zn}), one
can arrive at \be \label{33}
 Z_{g,n}(U_1,\dots,U_n;A) = \int D\psi(x)
 \exp{ (S_0[\psi] + S'[\psi] )},
 \ee
 where
 \begin{equation}\label{34}
S_0[\psi ] = -N^{2}\Omega{\Biggr (}2A \int _{0}^{1/2}W[ \psi (x)]
dx{\Biggl )} +2N^2(1-(g+s +r/2))\int_{0}^{1/2} dx \int_0^{1/2} dy
\log|\psi^2 (x)- \psi^2 (y)|.
\end{equation}
The phase structure of this term is equivalent to the phase
structure  of the following action,
\begin{equation}\label{36}
S'_0[\psi ] = -N ^2\hat{A} \int _{0}^{1/2} W[\psi (x)] dx
+2N^2(1-(g+s +r/2))\int_{0}^{1/2} dx \int_0^{1/2} dy \log|\psi^2
(x)- \psi^2 (y)|,
\end{equation}
where \be \hat{A} = 4A\Omega' ( 2A\int _{0}^{1/2} W[\psi (x)] dx
) \ee
 and
 \be
 S'[\psi(x)] = -N^2 V \int_0^{1\over 2} \psi^2(x) dx,
 \ee which coming from characters of the gauge group.
 Now we consider the special case $W[ \psi
 (x)]=\frac{1}{2}\psi^2(x)$, in this case the partition function
 (\ref{33}) can be rewritten as
\be \label{parmod}
 Z_{g,n}= \int D\psi(x)
 e^{[-N ^2(\hat{A}+V) \int _{0}^{1/2} \psi^{2} (x) dx
+2N^2(1-(g+s +r/2))
  \int_{0}^{1/2} dx \int_0^{1/2} dy \log|\psi^2
(x)- \psi^2 (y)|] },
 \ee
 It is seen that this partition function  is equal to the partition
function on a non-orientable surface with modified  area
$\hat{A}+V $, genus $g$, $r$ copies of projective plane, $S$
copies of Klein bottle and without boundaries. This model has
third order phase transition \cite{k4} on non-orientable surface
with boundary, $g=0, r=1, s=1$ and modified  area ($\hat{A} + V$).

\section{Conclusion}
It was shown that the free energy of the U$(N)$ YM$_2$ on a sphere
with the surface area $A<A_{\mathrm{c}}=\pi^2$ has a logarithmic
behavior \cite{15}. In \cite{dk}, the free energy was calculated
for areas $A>\pi^2$, from which it was shown that the YM$_2$ on a
sphere has a third-order phase transition at the critical area
$A_{\mathrm{c}}=\pi^2$. For surfaces with boundaries, the
situation is much more involved. In these cases, for each
boundary, the character of the holonomy of the gauge filed
corresponding to that boundary appears in the expression of the
partition function. If we denote the $j$'th boundary by $C_j$,
then each boundary condition is specified by the conjugacy class
of the holonomy matrix $U_j=$Pexp$\oint_{C_j}{\rm
d}x^\mu\;A_\mu(x)$. So, the boundary condition corresponding to
$C_j$ is fixed by the eigenvalues of $U_j$. These eigenvalues are
unimodular, so that each of them is specified by a real number
$\theta$ in $[-\pi,\pi]$. In the large-$N$ limit, the eigenvalues
of these matrices become continuous and one can denote the set of
these eigenvalues (corresponding to the $j$'th boundary) by an
eigenvalue density $\sigma_j(\theta),\theta\in [-\pi,\pi]$.
\\
In this paper we have studied the large-$N$ behavior of non-local
YM$_{2}$ on an orientable and non-orientable genus-$g$ surface
with $n$ boundaries ($\Sigma_{g,n}$). Here we have  restricted
ourselves to the cases in which the boundary holonomies $U_j$s
are close to identity. We have  shown that the critical behavior
of non-local YM$_2$ on $\Sigma_{g,n}$ with area $A(\Sigma_{g,n})$
is the same as a genus-$g$ surface with no boundary
($\Sigma_{g,0}$), but with the area
$A(\Sigma_{g,0})=A(\Sigma_{g,n})+V(U_1,\dots,U_n)$. Therefore, it
is seen that in the large-$N$ limit, the phase structure of
non-local YM$_2$ on $\Sigma_{g>0,n}$ is trivial, while non-local
YM$_2$ on $\Sigma_{0,n}$ exhibits a third-order phase transition.
Therefore the boundary conditions do not change the structure of
the phase transition.
  \vspace{3mm}
  \section{Acknowledgements}
I gratefully acknowledge helpful discussion with Prof. M.
Khorrami.
\section{Appendix}
It has been shown in \cite{KM} that in order to obtain the
Yang-tableau density corresponding to the dominant representation,
one should solve the generalized Hopf equation \be\label{14}
{\partial\over{\partial t}}(v\pm
i\pi\sigma)+{\partial\over{\partial\theta}} G[-i(v\pm
i\pi\sigma)]=0, \ee with the boundary conditions \bea\label{15}
\sigma(t=0,\theta)&=&\sigma_1(\theta)\cr
\sigma(t=\hat{A},\theta)&=&\sigma_2(\theta). \eea Then, if there
exists some $t_0$ for which \be\label{16} v(t_0,\sigma)=0, \ee one
denotes the value of $\sigma$ for $t=t_0$ by $\sigma_0$:
\be\label{17} \sigma_0(\theta):=\sigma(t_0,\theta), \ee and the
desired density satisfies \be\label{18}
\pi\rho[-\pi\sigma_0(\theta)]=\theta \ee What is shown is that
from this point of view, the non-local theory behaves like a
local theory but with a surface area $\hat{A}$ instead of $A$.
Note, however, that $\hat{A}$ itself depends on the Yang-tableau
density of the dominant representation, through \be\label{56}
\hat{A} = A \Omega'{\Biggr (}A \int _{0}^{1} W[\phi(x)] dx{\Biggl
)}. \ee
 or equivalently
 \be\label{58} \hat{A} =A \Omega'{\Biggr (}\int
dz\rho(z)W(z){\Biggl )}. \ee In \cite{15}, the critical area for a
Yang-Mills theory on a disk, $\sigma_1(\theta)=\delta(\theta)$,
has been found as: \be\label{21} A_{\rm
c}^{-1}={1\over\pi}\int{{d\theta'\;\sigma_2(\theta')}\over {\pi
-\theta'}}. \ee For a sphere, $\sigma_2(\theta)=\delta(\theta)$,
and one arrives at the familiar result \be\label{22} A_{\rm
c}=\pi^2. \ee These results can be used to obtain the critical
area for a non-local Yang-Mills theory on a disk. One can obtain
$\hat{A}_{\rm c}$ as \be\label{Ac} \hat{A}_{\rm
c}=({1\over\pi}\int{{d\theta'\;\sigma_2(\theta')}\over {\pi
-\theta'}})^{-1}. \ee To obtain $A_{\rm c}$ from $\hat{A}_{\rm
c}$, using (\ref{58}), one needs the critical density $\rho_{\rm
c}$. Even for the disk, it is not easy to find a closed form for
$\rho_{\rm c}$ for arbitrary $\sigma_2$.\\
In our problem, where, the holonomies corresponding to the
boundaries don't differ much from unity, $U\approx I$,
$\sigma(\theta)$ is nonnegligible only at $\theta$ near zero.
Expanding the denominator of (\ref{Ac}) up to second order in
$\theta$, one arrives at
\begin{equation}
\hat{A}_{\mathrm{c}}=\pi^2\left[\int\mathrm{d}\theta\;\sigma(\theta)
+\frac{1}{\pi}\int\mathrm{d}\theta\;\sigma(\theta)\,\theta
+\frac{1}{\pi^2}\int\mathrm{d}\theta\;\sigma(\theta)\,\theta^2+\cdots
\right]^{-1}
=\pi^2-\int\mathrm{d}\theta\;\sigma(\theta)\,\theta^2+\cdots.
\end{equation}
This is consistent with our general result(\ref{ac}) , since in
this case the second term of $V$ in (\ref{u}) vanishes:
\begin{equation}
V\approx \int\mathrm{d}\theta\;\sigma(\theta)\,\theta^2
\end{equation}

\end{document}